\RequirePackage{fix-cm}
\documentclass[smallextended]{svjour3}       
\smartqed  
\usepackage{graphicx}
\usepackage{amsmath}
\usepackage{algorithm}
\usepackage{algorithmic}
\usepackage[colorlinks=true,backref=page,bookmarksdepth=2]{hyperref}
\graphicspath{ {./figures/} }
\begin{document}

\title{Revisiting Lambert's Problem}


\author{Dario Izzo
}


\institute{D. Izzo \at
              ESTEC, Noordwijk, 2201 AZ \\
              Tel.: +123-45-678910\\
              Fax: +123-45-678910\\
              \email{dario.izzo@esa.int}           
}

\date{Received: date / Accepted: date}

\maketitle

\begin{abstract}
The orbital boundary value problem, also known as Lambert Problem, is revisited. Building upon Lancaster and Blanchard approach, new relations are revealed and a new variable representing all problem classes, under L-similarity, is used to express the time of flight equation. In the new variable, the time of flight curves have two oblique asymptotes and they mostly appear to be conveniently approximated by piecewise continuous lines. We use and invert such a simple approximation to provide an efficient initial guess to an Householder iterative method that is then able to converge, for the single revolution case, in only two iterations. The resulting algorithm is compared, for single and multiple revolutions, to Gooding's procedure revealing to be numerically as accurate, while having a significantly smaller computational complexity.

\end{abstract}

\section{Introduction}
\label{intro}
Lambert's problem, sometimes referred to as orbital boundary value problem, is a fascinating problem in astrodynamics that intrigued, over the years, most famous mathematicians. Just like Kepler's equation, its solution is at the very heart of fundamental astrodynamical and  space engineering questions \cite{celmech1,celmech2,izzolambert}. Following the fundamental work laid down, among others, by Euler, Lambert, Lagrange and Gauss, the need of having one robust algoritmic procedure able to function for a wide set of conditions led to revisit the Lambert's problem during the space era. Among the many contributions made during that period, the work of Lancaster and Blanchard \cite{lancaster} is to be highlighted as it reduced the solution to Lambert's problem to performing iterations each one requiring the computation of one only inverse trigonometric or hyperbolic function. Later, Gooding \cite{gooding} built upon these results and published a procedure achieving high precision in  only three iterations for all geometries. Gooding's algorithm makes use of Halley's iterations sided to well designed heuristics to set the initial guess of the iterated variable. His methodology to reconstruct the terminal velocity vectors is also remarkable as it is purely algebraic. The resulting procedure is extremely efficient having low computational cost and high accuracy. A number of studies \cite{peterson}, \cite{klumpp} and \cite{parrish} have tested Gooding approach extensively, suggesting its superiority with respect to other Lambert solvers. His procedure is most accurate and considered as the fastest existing approach to solve Lambert's problem \cite{arora}. Aside from Gooding's algorithm,  many other proposal have been put forward to design Lambert solvers, they all differ in the details of at least one of three fundamental ingredients: a) the iteration variable (directly connected to the time of flight equation), b) the iteration algorithm c) the initial guess and d) the reconstruction of the terminal velocity vectors. More recently  iimprovements on the original Gooding algorithm were also claimed \cite{arora} making use of the universal variable formulation \cite{bate} and an original cosine transformation. At the same time, a number of works recently addressed the possibility of deploying a large number of Lambert's algorithms on modern GPU architectures \cite{parrish}, \cite{arora2} and \cite{wie}. Interestingly, in the first of these works, a comparison is also made between Gooding procedure, a universal variable Lambert's solver and an early (slow) version of the algorithm here described (unpublished at that time) showing already its promising nature.

In this paper, we build upon Lancaster and Blanchard work, first deriving some new results, and then proposing and testing a new algorithm. The new algorithm a) iterates on the Lancaster-Blanchard variable $x$ using b) a Householder iteration scheme c) feeded by a simple initial guess found exploiting new analytical results found. The resulting procedure is simple to implement, does not make use of heuristics for the initial guess generation and is able to converge, on average, in only 2 iterations for the single revolution case and 3 in the multirevolution case, introducing a significant reduction in the overall solver complexity. 

\section{Background}
\label{sec:1}
\subsection{From Lambert to Gauss}
\label{sec:2}
Lambert's theorem states that the time of flight $t$ to travel along a keplerian orbit from $\mathbf r_1$ to $\mathbf r_2$ is a function of the orbit semi-major axis $a$, of the sum $r_1+r_2$ and of the chord $c$ of the triangle having $\mathbf r_1$ and $\mathbf r_2$ as sides. The complete formal proof was first delivered by Lagrange and is here sketched briefly in the form reported by Battin \cite{battin} as some of the quantities and equations involved will prove to be useful in our later developments. We start 
introducing the eccentric anomaly $E$ and the hyperbolic anomaly  $H$ via the corresponding Sundmann transformations $rdE = ndt$ $rdH = Ndt$. The mean motion $n = \sqrt{\mu / a^3}$ and its hyperbolic equivalent $N =  \sqrt{-\mu / a^3}$ are also introduced. As we do not make use of universal variables we will be forced to give all our arguments twice: one for the elliptic case $a>0$ and one for the hyperbolic case $a<0$. To this purpose some of the equations will be split in two lines, in which case the line above holds for the elliptic case and the line below holds in the hyperbolic case. We also make use of the reduced eccentric anomaly $E_r \in [0,2\pi]$ so that when $\tilde M$ full revolutions are made $E = E_r + 2\tilde M\pi$.  To ease the notation, in the following, we will drop the subscript $r$ so that $E$ will be the reduced eccentric anomaly.
The following relations are then valid for an elliptic orbit ($a > 0$):
\begin{equation}
\label{eq:ellipse}
\begin{array}{c}
r = a(1-e \cos E) \\
nt = E - e \sin E +2\tilde M\pi\\
r \cos f = a (\cos E - e) \\
r \sin f = a \sqrt{(1-e^2)} \sin E
\end{array}
\end{equation}
The first one relates the orbital radius $r$ to the eccentric anomaly $E$, the second one is the famous Kepler's equation relating the eccentric anomaly to the time of flight and the following two relation define the relations between true anomaly $f$ and eccentric anomaly $E$. Similar equations hold in the case of hyperbolic motion:
\begin{equation}
\label{eq:hyperbola}
\begin{array}{c}
r = a(1-e \cosh H) \\
Nt = e \sinh H - H\\
r \cos f = a (\cosh H - e) \\
r \sin f = -a \sqrt{(e^2-1)} \sinh H
\end{array}
\end{equation}
The above equations are valid along a Keplerian orbit, including $\mathbf r_1$ and $\mathbf r_2$. The time of flight can thus be written as:
\begin{equation}
\begin{array}{c}
\sqrt{\mu}(t_2-t_1)= \left\{ \begin{array}{l} a^{3/2} \left(E_2-E_1 + e\cos E_1 - e\cos E_2+2M\pi\right) \\ -a^{3/2} \left(e\cosh H_2 - e\cosh H_1 - (H_2 - H_1)\right) \end{array} \right.
\end{array}
\end{equation}
where $M = \tilde M_2 - \tilde M_1$ is the number of complete revolutions made during the transfer from $r_1$ to $r_2$. We may then define two new quantities such that:
\begin{equation}
\begin{array}{c}
\psi = \left\{ \begin{array}{l} \frac{E_2-E_1}2 \\ \frac{H_2-H_1}2 \end{array} \right., 
\cos\varphi = \left\{ \begin{array}{l} e\cos\frac{E_2+E_1}2 \\ e\cosh\frac{H_2+H_1}2 \end{array} \right.
\end{array}
\end{equation}
so that, by construction, in both the elliptic and hyperbolic motion case $\psi\in[0,\pi]$. We also restrict $\varphi\in[0,\pi]$ (elliptc case) and $\varphi\ge0$ (hyperbolic case) as to avoid ambiguity in the definition of the new angle.
The time of flight equation is then written as:
\begin{equation}
\label{eq:tof_psi}
\sqrt{\mu}(t_2-t_1)= \left\{ \begin{array}{l} 2 a^{3/2} \left(\psi - \cos\varphi\sin\psi + M\pi)\right. \\ -2a^{3/2} \left(\cosh\varphi\sinh\psi - \psi\right) \end{array} \right.
\end{equation}
The two new quantities introduced, $\varphi$ and $\psi$ only depend on the problem geometry and the sami-major axis $a$ as can be easily found by computing $c^2 = (r_2\cos f_2 - r_1\cos f_1)^2 - (r_2\sin f_2 - r_1\sin f_1)^2$ and $r_1 + r_2$ from Eq.(\ref{eq:ellipse}) and Eq.(\ref{eq:hyperbola}), holding:
\begin{equation}
\label{eq:g1}
r_1+r_2 = \left\{ \begin{array}{l} 2a(1-\cos\psi\cos\varphi) \\ 2a(1-\cosh\psi\cosh\varphi) \end{array} \right.
\end{equation}
\begin{equation}
\label{eq:g2}
c = \left\{ \begin{array}{l} 2a\sin\psi\sin\varphi \\ -2a\sinh\psi\sinh\varphi  \end{array} \right.
\end{equation}
Thus, one can conclude that the time of flight, given in Eq.(\ref{eq:tof_psi}), is a function of $a$, $c$ and $r_1+r_2$. To further investigate the functional relation of the time of flight to these quantities it is convenient to introduce two new angles:
\begin{equation}
\alpha = \varphi+\psi, \beta = \varphi-\psi
\end{equation}
Clearly, in the elliptic case $\alpha\in[0.2\pi]$ and $\beta\in[-\pi,\pi]$ while for the hyperbolic case $\alpha\ge0$ and $\beta\ge-\pi$. These bounds are very important, as we shall see, in solving a quadrant ambiguity of the newly defined quantities. The time of flight equation now takes the elegant form:
\begin{equation}
\label{eq:tof_alpha}
\sqrt{\mu}(t_2-t_1)= \left\{ \begin{array}{l} a^{3/2} \left(\left(\alpha-\sin\alpha)-(\beta-\sin\beta\right) + 2M\pi\right) \\ -2a^{3/2} \left(\left(\sinh\alpha-\alpha\right)-\left(\sinh\beta-\beta\right)\right) \end{array} \right.
\end{equation}
and computing $r_1+r_2 \pm c$ from Eq.(\ref{eq:g1}) and Eq.(\ref{eq:g2}) one easily finds:
\begin{equation}
\label{eq:g5}
\frac s{2a} = \left\{ \begin{array}{l} \sin^2\frac\alpha 2 \\ -\sinh^2\frac\alpha 2 \end{array} \right.
\end{equation}
\begin{equation}
\label{eq:g6}
\frac {s-c}{2a} = \left\{ \begin{array}{l} \sin^2\frac\beta 2 \\ -\sinh^2\frac\beta 2 \end{array} \right.
\end{equation}
These last three equations were first derived by Lagrange and used in his proof of the Lambert's theorem. The angles $\alpha$ and $\beta$ cannot be determined univoquely from the equations above as their quadrant is not defined. We thus appear to have two possible solutions for $\alpha$ and $\beta$. The quadrant of $\beta$ can actually be resolved by expanding $\cos \theta /2 = \cos (f_2-f_1)/2$ using trigonometric identities and eventaully showing that the following holds:
\begin{equation}
\label{eq:cost2}
\sqrt{r_1r_2}\cos\frac\theta 2 = 
\left\{ 
\begin{array}{l} 
2a\sin\frac\alpha 2\sin \frac\beta 2 \\
-2a\sinh\frac\alpha 2\sinh \frac\beta 2
\end{array} 
\right.
\end{equation}
since $\sin\frac\alpha 2,\sinh\frac\alpha 2 \ge 0$ the above equations dictate that $\sin\frac \beta 2, \sinh\frac \beta 2$ have the same sign as $\cos\frac\theta 2$, thus $\beta \in [-\pi,0]$ when $\theta \ge \pi$ and $\beta > 0$ when $\theta \in[0,\pi]$. The ambiguity on the $\alpha$ angle, instead, cannot be resolved as it derives from the fact that exactly two different ellipses, having the same semi-major axis $a$, link $\mathbf r_1$ and $\mathbf r_2$ and thus two different time of flights exist that satisfy Eq.(\ref{eq:tof_alpha}). From Eq.(\ref{eq:g5}) and Eq.(\ref{eq:g6}) one can also derive the useful relation:
\begin{equation}
\label{eq:g7}
\sin\frac\alpha 2 = \lambda\sin\frac\beta 2
\end{equation}

\subsection{The Lambert's problem revival}
During the 18th-19th century, the work on the orbital boundary value problem culminated with Gauss masterpiece \lq\lq Theoria Motus Corporum Coelestium in Sectionibus Conicis Solem Ambientium\rq\rq\ \cite{gauss} where the \lq\lq prince of mathematicians\rq\rq\ conceives was is probably the first procedure able to accurately solve the Lambert's problem (see Battin \cite{battin} for an excellent account of Gauss method). In the following years science drifted slowly away this topic, only to revisit it in the second half of the 20th century when the orbital boundary value problem received more attention in the context of Moon exploration. Hence, the work of Lancaster, Blanchard, Battin, Bate and many others introduced several advances on the topic. We here follow the approach from Lancaster and Blanchard that inspred most of our developments and we will thus rederive some of their relations which are needed to explain our new ideas. Consider the parameter $\lambda$ defined as:
$$
s \lambda = \sqrt{r_1r_2}\cos\frac\theta 2
$$
using Eq.(\ref{eq:cost2}) and substituting the expressions in  Eq.(\ref{eq:g5}) and Eq.(\ref{eq:g6}) it is simple to show that:
$$
\lambda^2 = \frac{s-c} s
$$
The parameter $\lambda \in [-1,1]$ is positive when $\theta \in[0,\pi]$ and negative when $\theta\in[\pi,2\pi]$. Values of $\lambda^2$ close to unity indicate a chord of zero length, a case which is indeed extremely interesting in interplanetary trajectory design as it is linked to the design of resonant transfers. We also introduce a non dimensional time-of-flight defined as:
$$
T = \frac 12 \sqrt{\frac \mu{a_m^3}} (t_2-t_1) = \sqrt{2\frac \mu{s^3}} (t_2-t_1)
$$
where $a_m = s/2$ is the minimum energy ellipse semi-major axis \cite{battin}. The advantage of using $\lambda$ and $T$ derives from the fact that $T$ is a function of $a / a_m$ and $\lambda$ alone, which allows to greatly simplify the taxonomy of possible Lambert's problems. In Gooding's words \cite{gooding}, all the triangles having equal $c / s$ ratio form a large equivalence class and can be described as L-similar. For them, all Lambert solutions are the same in terms of $a / a_m$ and $T$. 

\begin{figure}
  \includegraphics[width=0.75\textwidth]{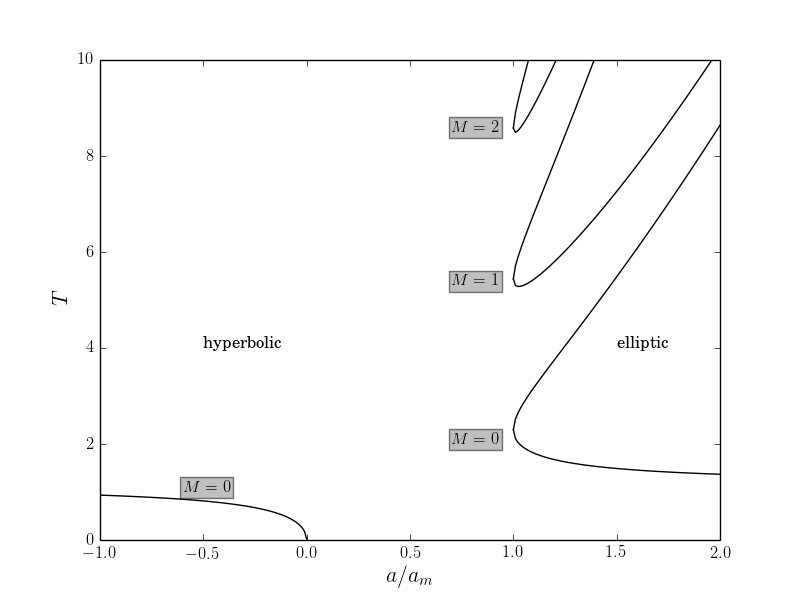}
\caption{Non-dimensional time of flight curve for $\lambda=-0.9$ parametrized using $a/a_m$ . \label{fig:tof_a}}
\end{figure}

If we now plot the time of flight given by Eq.(\ref{eq:tof_alpha}) as a function of the ratio between the semi-major axis and the minimum energy ellipse semi-major axis, for a particular value of $\lambda$ and for single and multiple revolution cases, we get Figure \ref{fig:tof_a}. It is evident how, in order to invert the time of flight relation iterating over $a / a_m$, while possible, is not a good choice. To avoid these problems we follow Lancaster and Blanchard in some further derivations introducing the new quantities:
\begin{equation}
\label{eq:def_x}
\begin{array}{cc}
x = \left\{ \begin{array}{l} \cos\frac\alpha 2 \\ \cosh\frac\alpha 2 \end{array}\right., &
y = \left\{ \begin{array}{l} \cos\frac\beta 2 \\ \cosh\frac\beta 2 \end{array}\right.
\end{array}
\end{equation}
which imply:
\begin{equation}
\label{eq:def_x}
\begin{array}{cc}
\left\{ \begin{array}{l} \sqrt{1-x^2} = \sin\frac\alpha 2  \\ \sqrt{x^2-1} = \sinh\frac\alpha 2  \end{array}\right., &
\left\{ \begin{array}{l} \lambda \sqrt{1-x^2} = \sin\frac\beta 2 \\ \lambda \sqrt{x^2-1} = \sinh\frac\beta 2  \end{array}\right.
\end{array}
\end{equation}
and $y = \sqrt{1-\lambda^2(1-x^2)}$. Using these relations it is possible to relate the auxiliary angles $\varphi$ and $\psi$ directly to $x$:
\begin{equation}
\label{eq:g8}
\begin{array}{l} 
\cos\varphi = xy - \lambda(1-x^2) \\
\cosh\varphi = xy + \lambda(x^2-1)
\end{array} 
, \hskip1cm
\begin{array}{l} 
\sin\varphi = (y+x\lambda)\sqrt{1-x^2} \\
\sinh\varphi = (y+x\lambda)\sqrt{x^2-1} 
\end{array} 
\end{equation}
and,
\begin{equation}
\label{eq:g9}
\begin{array}{l} 
\cos\psi = xy + \lambda(1-x^2) \\
\cosh\psi = xy - \lambda(x^2-1)
\end{array} 
, \hskip1cm
\begin{array}{l} 
\sin\psi = (y - x\lambda)\sqrt{1-x^2} \\
\sinh\psi = (y - x\lambda)\sqrt{x^2-1} 
\end{array} 
\end{equation}
which allows to derive the relations $\cos\varphi\sin\psi = (x-\lambda y)\sqrt{1-x^2}$, $\cosh\varphi\sinh\psi = (x-\lambda y)\sqrt{x^2-1}$ and thus have the following time of flight equation valid in all cases:
\begin{equation}
\label{eq:tof_x}
T = \frac{1}{1-x^2}\left( \frac{\psi+M\pi}{\sqrt{|1-x^2|}} - x + \lambda y\right)
\end{equation}
where we must set $M=0$ in the case of hyperbolic motion where unbounded motion prevents complete revolutions to happen. The auxiliary angle $\psi$ is computed using Eq.(\ref{eq:g9}) by the appropriate inverse function and thus, the time of flight evaluation is reduced to one only inverse function computation. Given the bounds on $\alpha$, from the definition of $x$, we can see how $x \in [-1,\infty]$. Also, $x>1$ implies hyperbolic motion, while $x<1$ elliptic motion. Since $1-x^2 = \sin^2\frac\alpha 2 = \frac s{2a} = \frac{a_m}a$, we see how $x=0$ corresponds to the minimum energy ellipse. Note that different Lambert's problems having identical $\lambda$ values (i.e. same $c / s$), result in the same $x$, we then say that $x$ is a Lambert invariant parameter. 

Computing Eq.(\ref{eq:tof_x}) in $x=0$ we get:
\begin{equation}
\label{eq:t0}
T(x=0)=T_{0M} = \arccos\lambda + \lambda\sqrt{1-\lambda^2} + M\pi = T_{00} + M\pi
\end{equation}
where we have introduced $T_0$ as the value of $T$ in $x=0$ and $T_{00}$ as the value in the single revolution case $M=0$.

\begin{figure}
  \includegraphics[width=0.75\textwidth]{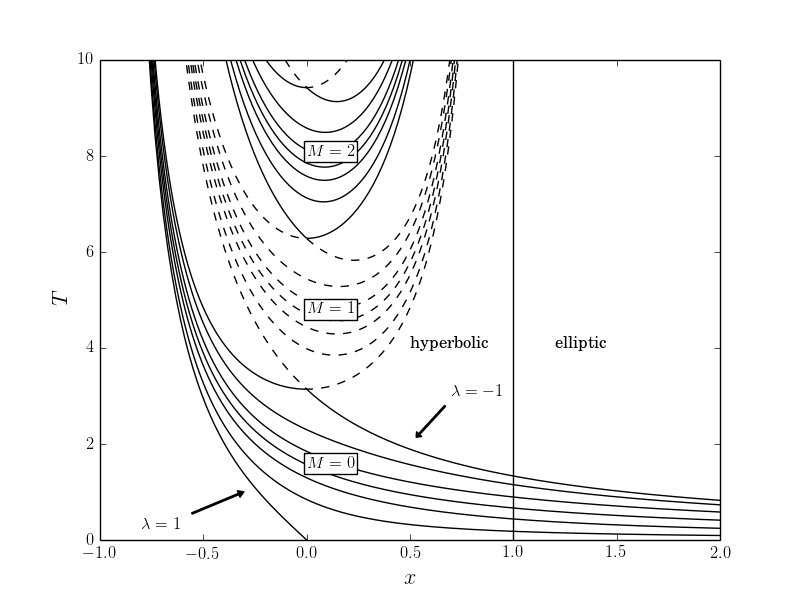}
\caption{Time of flight curves parametrized using $x$ for different $\lambda$ and $M$ values. \label{fig:tof_x}}
\end{figure}

When computing Eq.(\ref{eq:tof_x}) in the single revolution case, a loss of precision is encountered due to numerical cancellation for values of $x\approx 1$ where both $1-x^2$ and $\psi$ tend to zero. In these cases we compute the time of flight equation by series expansion using the elegant result from Battin \cite{battin} setting:
\begin{equation}
\label{eq:tof_x_series}
\begin{array}{l}
\eta = y - \lambda x \\
S_1 = \frac 12 (1-\lambda - x\eta) \\
Q = \frac 43 {}_1F_2(3,1,\frac 52,S_1) \\
2T = \eta^3Q + 4\lambda\eta
\end{array}
\end{equation}
where ${}_1F_2(a,b,c,d)$ is the Gaussian or ordinary hypergeometric function. This can be evaluated by direct computation of the associated hypergeometric series. Noting that $S_1\to 0$ when $x\to 1$ the number of terms to retain in the series is small whenever the series is used in the neighbourhood of $x = 1$. Departing from Battin, we study the parabolic case substituting $x=1, y=1$ into Eq.(\ref{eq:tof_x_series}) and thus obtaining the following remarkable expression:
\begin{equation}
\label{eq:t1}
T(x=1) = T_1 = \frac{2}{3}(1-\lambda^3)
\end{equation}
relating the geometry of the triangle created by two different observations of an object on a parabolic keplerian orbit to the non-dimensional time elapsed between them. It is also possible to derive the following formulas for the time of flight derivatives:
\begin{equation}
\label{eq:derivatives}
\begin{array}{l}
(1-x^2)\frac {dT}{dx} = 3Tx-2+2\lambda^3\frac xy\\
(1-x^2)\frac {d^2T}{dx^2} = 3T + 5x\frac{dT}{dx}+2(1-\lambda^2)\frac{\lambda^3}{y^3}\\
(1-x^2)\frac{d^3T}{dx^3} = 7x\frac{d^2T}{dx^2} + 8 \frac{dT}{dx} - 6(1-\lambda^2)\lambda^5\frac x{y^5}
\end{array}
\end{equation}
which are valid in all cases (single and multiple revolutions, elliptic and hyperbolic) except in $\lambda^2=1, x=0$ and $x=1,\forall \lambda$. We then apply de l'H\^{o}pital rule to the first of the above equations, and using the expression derived for $T_1$ we are also able to find the value of the derivative of the time of flight curves in the case of a parabola:
\begin{equation}
\label{eq:dt1dx}
\left. \frac{dT}{dx}\right|_{x=1} = \frac 25 (\lambda^5-1)
\end{equation}
which is valid for $M=0$. By direct substitution, one can also easily show:
$$
\left. \frac{dT}{dx}\right|_{x=0} = -2
$$
A great advantage of the time of flight equation in the form of Eq.(\ref{eq:tof_x}) as derived by Lancaster and Blanchard \cite{lancaster} is in the low computational cost of computing $T$ and its derivatives, up to the third order. Only one trigonometric (or hyperbolic) function inversion, two square roots and a few multiplications, divisions and sums are indeed necessary to compute these numerical values. Other approaches based on geomerical considerations or on a universal variables formulation are, at best, only able to match such a simple representation. We now summarize all the information relative to all possible Lambert problems in one single graph as done in Figure \ref{fig:tof_x}.
\section{A new Lambert invariant variable}
\begin{figure}
\includegraphics[width=0.49\textwidth]{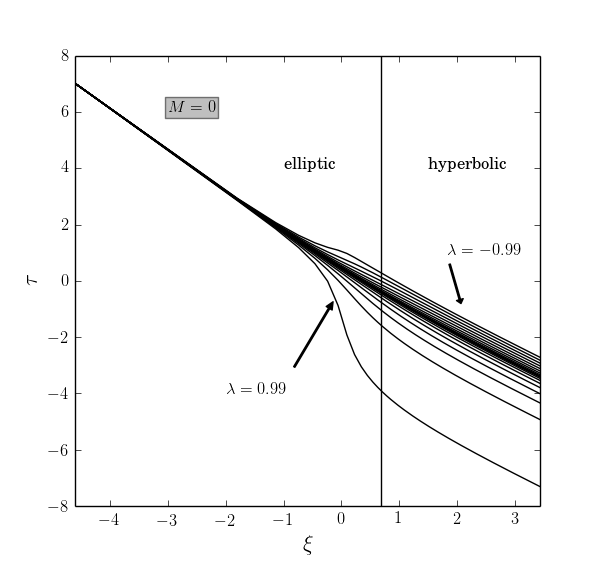}
\includegraphics[width=0.49\textwidth]{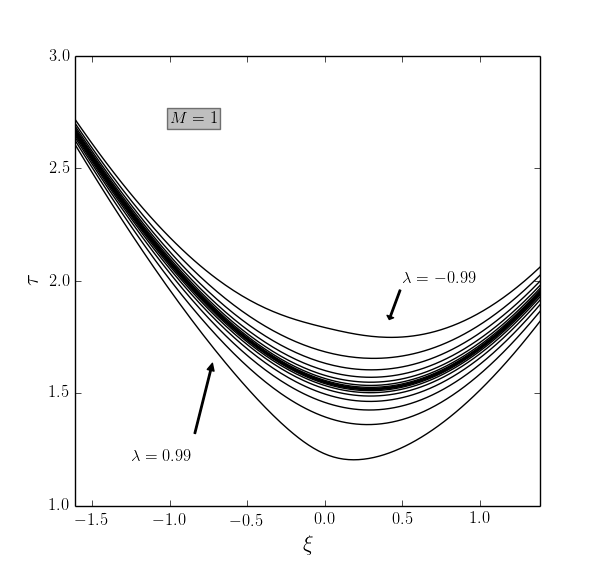}
\caption{Time of flight curves ($\tau$) parametrized using $\xi$ for 30 $\lambda$ values equally spaced in $[-0.9,0.9]$. \label{fig:tof_xi}}
\end{figure}

Let us consider the following new variables:
\begin{equation}
\label{eq:new_variables}
\xi = \left\{ 
\begin{array}{ll}
\log(1+x), &M=0 \\
\log(\frac{1+x}{1-x}), &M>0
\end{array} \right., \hskip1cm
\tau = \log(T)
\end{equation}
The domain of the time-of flight curve is now extended to $[-\infty,\infty]$. In the case of $M=0$ the co-domain is also extended similarly. Let us study the resulting time of flight equation $\tau(\xi,\lambda,M)$. In Figure \ref{fig:tof_xi} we plot $\tau$ against $\xi$ for $M=0$ and $M=1$ and thirty equally spaced values of $\lambda\in[-0.9,0.9]$. In the case $M=0$ the curves appear to have two asymptotes having negative inclination coefficient. For the multiple revolution case ($M>0$) the curves have two symmetric asymptotes. The new introduced parameter $\xi$ is Lambert invariant according to Gooding's definition \cite{gooding} as it essentially is a transformation of the Lambert invariant variable $x$. We study the differential properties of the new curves, we have:

\begin{equation}
\label{eq:new_variables_differentials}
d\xi = \left\{ 
\begin{array}{ll}
\frac{1}{1+x}dx, &M=0 \\
\frac{2}{1-x^2}dx, &M>0
\end{array} \right., \hskip1cm
d\tau = \frac 1T dT
\end{equation}
Substituting these relations into Eq.(\ref{eq:derivatives}), after some manipulations we may derive the following hybrid expressions for the derivatives in the case $M=0$:
\begin{equation}
\label{eq:derivatives_tau}
\begin{array}{l}
\frac {d\tau}{d\xi} = \frac{1+x}T\frac{dT}{dx}\\
\frac {d^2\tau}{d\xi^2} = \frac{(x+1)^2}{T}\frac{d^2T}{dx^2}+\frac{d\tau}{d\xi}-\left(\frac{d\tau}{d\xi}\right)^2\\
\frac{d^3\tau}{d\xi^3} = \frac{(1+x)^3}{T}\frac{d^3T}{dx^3} + \left(\frac{d^2\tau}{d\xi^2}-\frac{d\tau}{d\xi}+\left(\frac{d\tau}{d\xi}\right)^2\right)\left(2-\frac{d\tau}{d\xi}\right)+\frac{d^2\tau}{d\xi^2} - 2\frac{d\tau}{d\xi}\frac{d^2\tau}{d\xi^2}
\end{array}
\end{equation}
Note that these expressions can be computed, sequentially, after Eq.(\ref{eq:derivatives}). The following holds for the $M=0$ case:
\begin{equation}
\label{eq:asympt1}
\begin{array}{l}
\lim_{\xi\to\infty} \tau = -\xi+\log(1-\lambda|\lambda|)\\
\lim_{\xi\to-\infty} \tau = -\frac 32 \xi + \log\left(\frac{\pi}{4}\sqrt{2}\right)
\end{array}
\end{equation}
which describes the asymptotic behaviour of the time of flight as visualized in Figure \ref{fig:tof_xi}. The two asymptotes are thus revealed to have negative coefficients $-1$ and $-3 / 2$. For the multirevolution cases the derivatives are found to be:
\begin{equation}
\label{eq:derivatives_tau_mr}
\begin{array}{l}
\frac {d\tau}{d\xi} = \frac{1-x^2}{2T}\frac{dT}{dx}\\
\frac{d^2\tau}{d\xi^2}=\frac{(1-x^2)^2}{4T}\frac{d^2T}{dx^2} - x\frac{d\tau}{d\xi} - \left(\frac{d\tau}{d\xi}\right)^2\\
\begin{split}
\frac{d^3\tau}{d\xi^3}=\frac{(1-x^2)^3}{8T}\frac{d^3T}{dx^3}-\left(\frac{d^2\tau}{d\xi^2}+x\frac{d\tau}{d\xi}+\left(\frac{d\tau}{d\xi}\right)^2\right)\left(2x+\frac{d\tau}{d\xi}\right) - \\
-\frac{1-x^2}{2}\frac{d\tau}{d\xi}-x\frac{d^2\tau}{\xi^2}-2\frac{d\tau}{d\xi}\frac{d^2\tau}{d\xi^2}
\end{split}
\end{array}
\end{equation}
again computable in cascade and the following asymptotic behavior can be derived:
\begin{equation}
\label{eq:asympt2}
\begin{array}{l}
\lim_{\xi\to-\infty} \tau = \log\left(\frac{\pi+M\pi}8\right) - \frac 32 \xi\\
\lim_{\xi\to\infty} \tau = \log\left(\frac{M\pi}8\right) + \frac 32 \xi
\end{array}
\end{equation}
revealing two symmetric asymptotes having inclination $\pm 3 / 2$.
\begin{figure}
\includegraphics[width=0.95\textwidth]{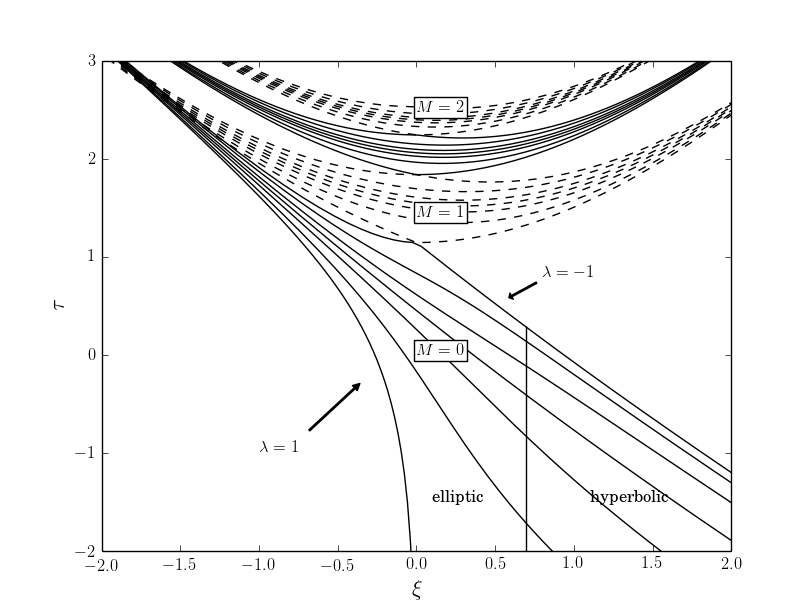}
\caption{Time of flight curves parametrized using $\xi$ for different $\lambda$ and $M$ values. \label{fig:tof_xi_tot}}
\end{figure}
In Figure \ref{fig:tof_xi_tot} we report the time of flight curves in the new variables for different values of $\lambda$ and $M$. The reader can then compare this $\xi$-$\tau$ plane to the $x$-$T$ plane visualised in Figure \ref{fig:tof_x}. 

\section{Lambert Solver}

A Lambert solver can be defined as a procedure that returns, for a gravitational field of strength $\mu$, all the possible velocity vectors $\mathbf v_1$ and $\mathbf v_2$ along keplerian orbits linking $\mathbf r_1$, $\mathbf r_2$ in a transfer time $T^*$. The ingredient of such an algorithm are, essentially a) the choice of a variable to iterate upon and thus invert the time of flight curve, b) the iteration method, c) the starting guess to use with the iteration method and d) the reconstruction methodology to compute $\mathbf v_1$ and $\mathbf v_2$ from the value returned by the iterations. As we will detail, our Lambert solver a) iterates on the Lancaster-Blanchard variable $x$ using b) a Householder iteration scheme c) feeded by initial guesses found exploiting the curve shape in the $\tau$-$\xi$ plane and the  new analytical results found above. Eventually the velocity vectors are reconstructed following, again, the methodology proposed by Gooding \cite{gooding}. The final pseudo-code of the proposed Lambert solver is reported in Algorithm \ref{alg:main}, \ref{alg:findxy}. Note how we detect the maximum number of revolutions $M_{max}$ at the beginning by computing $T_{min}$ in one case. All other cases (i.e. $M<=M_{max}$) will not require the evaluation of a $T_{min}$ via an iterative procedure. By doing so we do not bound the roots (short and long period) and thus risk cases where the solution jumps between the long and short period branches. While this did not happen in our tests of the new routine, it is a possibility we are not safeguarding against. 

The code, written in C++ and exposed to python, is made available as part of the open source project PyKEP from the European Space Agency github repository \url{https://github.com/esa/pykep/}. The final algorithm is the final result of many different trials to exploit the newly found results detailed above, and in particular the $T_0,T_1$ expressions (and their derivatives) and the $\xi$-$\tau$ plane. It is worth reporting how one very robust  set-up, not selected as our final proposed algorithm, was that of iterating directly with a simple derivative free method (regula-falsi) on the $\xi$-$\tau$ plane using constant initial guesses (i.e $x_l = -0.7, x_r=0.7$). We ended up choosing a different set-up (Algorithm \ref{alg:main}, \ref{alg:findxy}) which turned up to be faster in our computational tests.

\subsection{The Householder iterations}

One of the main differences of the proposed Lamber solver with respect to previous work is the use of the Householder iterative scheme as a root finder for the time-of-flight curves $T(x)-T^* =  0$. Householder iterations are not used widely as the added computational effort of computing higher order derivatives is not worth the gain whenever these request further function evaluations. In our case, as the derivatives computation is done using equations \ref{eq:derivatives}, Householder iterations are able to provide a significant benefit. We report the exact form used to implement the iterations as it is known how different numerical form can produce different behaviours. After experimenting with different implementations the following was used:

$$
x_{n+1} = x_n - f(x_n) \frac{f'^2(x_n) - f(x_n)f''(x_n) / 2}{f'(x_n) (f'^2(x_n) - f(x_n)f''(x_n)) + f'''(x_n)f^2(x_n) / 6 }
$$
where $f$ is, in our case, $T(x)-T^*$ and the derivatives are indicated as $f',f'',f'''$.

\subsection{Initial Guess}

To generate an initial guess for $x$ we use the newly introduced $\xi$-$\tau$ variables and the values $T_0$ and $T_1$ as computed from Eq.(\ref{eq:t0}) and Eq,(\ref{eq:t1}). Our initial guess is obtained inverting the following linear approxmation to the time of flight curves:
$$
\tau = c\xi + d
$$
where we vary the $c$ and $d$ values according to the value of $\tau$ and $M$.

\subsubsection{Single revolution}

Let us start form the single revolution case. Clearly, for high values of $\tau$, we must set $c = -1.0$, while for low values $c=-3/2$ so that the asymptotic behavior derived in Eq.(\ref{eq:asympt1}) is reproduced. We then consider the following piece-wise linear approximation:
$$
\begin{array}{ll}
\tau = -\frac 32 \xi + \tau_0, & T \ge T_0 \\
\tau = - \xi + \log 2 +\tau_1, & T < T_1 \\
\tau = \tau_0 + \frac{\xi}{\log 2}(\tau_1 - \tau_0) & T_1 < T < T_0
\end{array}
$$
where $\tau_1=\log T_1$ and $\tau_0$ = $\log T_0$. We have basically enforced the lines to pass through the points $(x_0,T_0)$ and $(x_1,T_1)$ and have the desired asymptotic behaviour.

\begin{figure}
\includegraphics[width=0.45\textwidth]{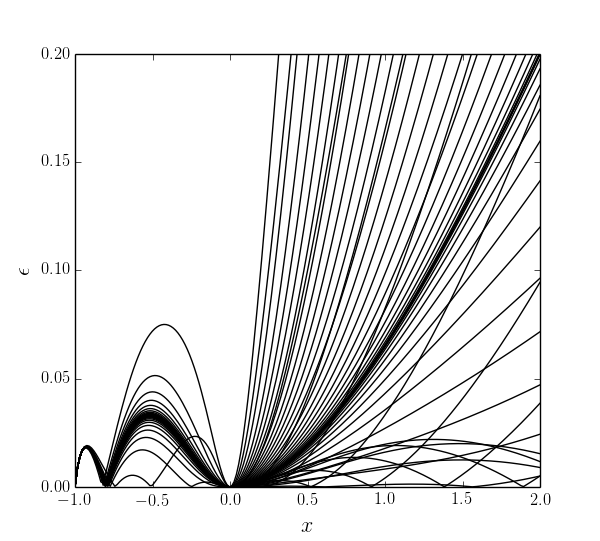}
\includegraphics[width=0.45\textwidth]{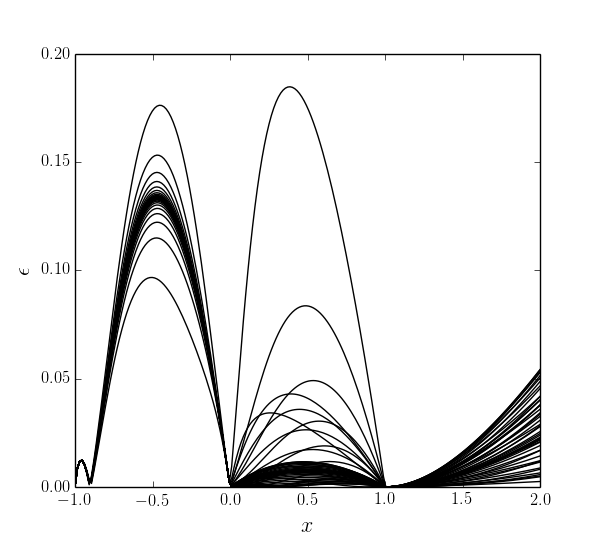}
\caption{Absolute errors introduced by the Gooding initial guess (left) and the proposed initial guess (right) for the single revolution $M=0$ case. Each line correspond to a different $\lambda$ value ranging from -0.99 to 0.99 \label{fig:errors}}
\end{figure}

The above relation is easily inverted and thus the following simple starter $\xi_0$ is derived:
$$
\begin{array}{ll}
\xi_0 = \frac 23(\tau_0-\tau), & T \ge T_0 \\
\xi_0 = \log 2+\tau_1-\tau, & T < T_1 \\
\xi_0 = \frac{\tau-\tau_0}{\tau_1-\tau_0} \log2 & T_1 < T < T_0
\end{array}
$$
Transforming these relations back to the $x$-$T$ plane we find the following expression for the starter $x_0$:
\begin{equation}
\label{eq:x0}
\begin{array}{ll}
x_0 = \left(\frac{T_0}{T}\right)^{\frac 23} -1, & T \ge T_0 \\ \\
x_0 = 2\frac{T_1}{T} -1, & T < T_1 \\
x_0 =  \left(\frac{T_0}{T}\right)^{\log_2\left(\frac {T_1}{T_0}\right)} -1& T_1 < T < T_0
\end{array}
\end{equation}
having an extremely low computational cost also in view of the fact $\log_2$ admits efficient implementaions. We can improve the expression for the $T<T_1$ case making use of the newly found result expressed in Eq.(\ref{eq:dt1dx}). We thus set 
$$
\begin{array}{ll}
x_0 = \left(\frac{T_0}{T}\right)^{\frac 23} -1, & T \ge T_0 \\ \\
x_0 = \frac 52 \frac{T_1(T_1-T)}{T (1-\lambda^5) } + 1, & T < T_1 \\
x_0 =  \left(\frac{T_0}{T}\right)^{\log_2\left(\frac {T_1}{T_0}\right)} -1& T_1 < T < T_0
\end{array}
$$
where we enforced the derivatives and values at $x=1$ and $x=\infty$. We report the error introduced by using these expressions in Figure \ref{fig:errors} where we also show, for comparison, the same plot relative to the Gooding initial guess. The error is defined by the difference between the initial guess computed for a given $T^*$ and the actual value of $x$ resulting in a time of flight $T^*$.

\subsubsection{Multiple revolutions}
For the multiple revolution case, assuming a solution exists, there are two possible values of $x$ and thus we will need two starters. We obtain these by direct inversion of Eq.(\ref{eq:asympt2}) that define the two asymptotes.
$$
\begin{array}{l}
\xi_{0l} = \frac 23(\log{\frac{\pi+M\pi}8}-\tau) \\
\xi_{0r} = \frac 23(\tau-\log{\frac{M\pi}8})
\end{array}
$$
The above equations may then be transformed back to the $x$-$T$ plane so that the following simple starters are derived:
\begin{equation}
\label{eq:x0lr}
\begin{array}{l}
x_{0l} = \frac{\left(\frac{M\pi + \pi}{8T}\right)^{\frac 23}-1}{\left(\frac{M\pi + \pi}{8T}\right)^{\frac 23}+1} \\ 
x_{0r} = \frac{\left(\frac{8T}{M\pi}\right)^{\frac 23}-1}{\left(\frac{8T}{M\pi}\right)^{\frac 23}+1} 
\end{array}
\end{equation}
The above expressions approximate well the time of flight curves as $|x| \to 1$. A great advantage of these expressions is that they do not make use of $T_{min}$, $x_{min}$ (i.e. the minimum of the time of flight curve and its extremal value) which would require a distinct set of iterations to be found. We then avoid to pass to the root solver solution bounds at each $M$, at the risk of allowing, during the iterations, switches between short and long period solutions, a theoretical occurence, though, that was never encountered in our experimental tests.

\begin{algorithm}[t]
\caption{Lambert solver: inputs, $\mathbf r_1=[r_{11},r_{12},r_{13}]$, $\mathbf r_2=[r_{21},r_{22},r_{23}]$, $t$ and $\mu$\label{alg:main}}
\begin{algorithmic}
\REQUIRE $t > 0$, $\mu > 0$
\STATE $\mathbf c = \mathbf r_2 - \mathbf r_1$
\STATE $c=|\mathbf c|$, $r_1=|\mathbf r_1|$, $r_2=|\mathbf r_2|$
\STATE $s=\frac 12 (r_1+r_2+c)$
\STATE $\hat {\mathbf i}_{r,1} = \mathbf r_1 / r_1$, $\hat {\mathbf i}_{r,2} = \mathbf r_2/ r_2$
\STATE $\hat {\mathbf i}_{h} = \hat {\mathbf i}_{r,1} \times \hat {\mathbf i}_{r,2}$
\STATE $\lambda^2 = 1-c/s$, $\lambda = \sqrt{\lambda^2}$
\IF {($r_{11}r_{22} - r_{12}r_{21}) < 0$}
\STATE $\lambda = -\lambda$
\STATE $\hat {\mathbf i}_{t,1} = \hat {\mathbf i}_{r,1} \times \hat{\mathbf i}_h  $, $\hat {\mathbf i}_{t,2} =  \hat {\mathbf i}_{r,2} \times \hat
{\mathbf i}_{r,2}$
\ELSE
\STATE $\hat {\mathbf i}_{t,1} = \hat{\mathbf i}_h \times \hat {\mathbf i}_{r,1}$, $\hat {\mathbf i}_{t,2} = \hat{\mathbf i}_h \times \hat 
{\mathbf i}_h$
\ENDIF
\STATE $T = \sqrt{\frac{2\mu}{s^3}} t$
\STATE $x_{list},y_{list} = $findxy($\lambda$, $T$)
\STATE $\gamma = \sqrt{\frac{\mu s}{2}}$, $\rho = \frac{r_1-r_2}{c}$, $\sigma = \sqrt{(1-\rho^2)}$
\FOR {each $x,y$ in $x_{list},y_{list}$}
\STATE  $V_{r,1} = \gamma [(\lambda y - x) - \rho(\lambda y + x)] / r_1$
\STATE  $V_{r,2} = -\gamma [(\lambda y - x) + \rho(\lambda y + x)] / r_2$
\STATE  $V_{t,1} = \gamma\sigma(y+\lambda x) / r_1$
\STATE  $V_{t,2} = \gamma\sigma(y+\lambda x) / r_2$
\STATE $\mathbf v_1 = V_{r,1} \hat {\mathbf i}_{r,1} + V_{t,1} \hat {\mathbf i}_{t,1}$
\STATE $\mathbf v_2 = V_{r,2} \hat {\mathbf i}_{r,2} + V_{t,2} \hat {\mathbf i}_{t,2}$
\ENDFOR
\end{algorithmic}
\end{algorithm}

\begin{algorithm}
\caption{findxy($\lambda$, $T$): computes all $x,y$ for single and muti-rev solutions \label{alg:findxy}}
\begin{algorithmic}
\REQUIRE $|\lambda| < 1$, $T < 0$
\STATE $M_{max} = \mbox{floor} (T / \pi)$
\STATE $T_{00} = \arccos \lambda + \lambda\sqrt{1-\lambda^2}$
\IF{$T < T_{00} + M_{max}\pi$ and  $M_{max}>0$}
\STATE start Halley iterations from $x=0,T=T_0$ and find $T_{min}(M_{max})$
\IF {$T_{min} > T$}
\STATE $M_{max} = M_{max} - 1$
\ENDIF
\ENDIF
\STATE $T_1 = \frac 23 (1-\lambda^3)$
\STATE compute $x_0$ from Eq.(\ref{eq:x0})
\STATE start Householder iterations from $x_0$ and find $x,y$
\WHILE{$M_{max}>0$}
\STATE compute $x_{0l}$ and $x_{0r}$ from Eq.(\ref{eq:x0lr}) with $M=M_{max}$
\STATE start Householder iterations from $x_{0l}$ and find $x_r,y_r$
\STATE start Householder iterations from $x_{0r}$ and find $x_l,y_l$
\STATE $M_{max}=M_{max}-1$
\ENDWHILE
\end{algorithmic}
\end{algorithm}

\section{New solver performances}
\begin{figure}
\includegraphics[width=0.95\textwidth]{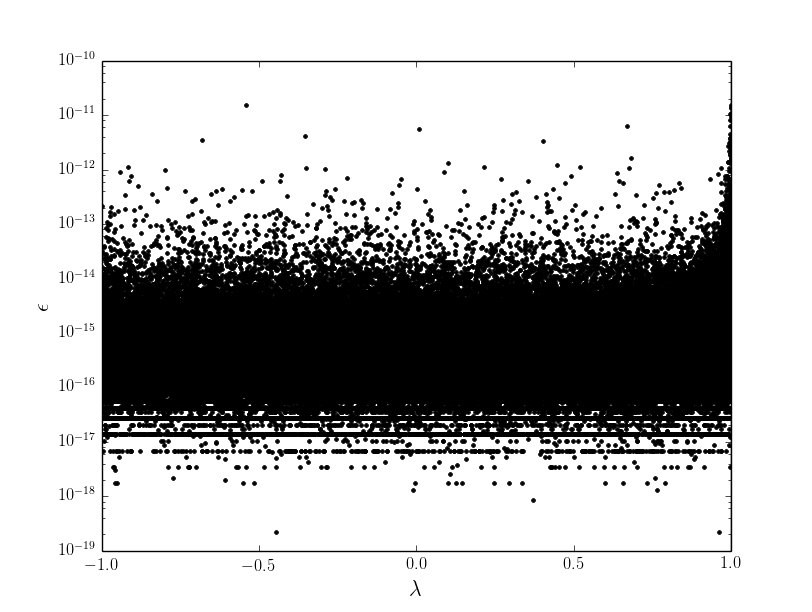}
\caption{Absolute error $\epsilon = |x-x_{true}|$ of our Lambert solver as a function of $\lambda$. This is achieved, on average over the $M=0$ cases, in 2 iterations \label{fig:accuracy}}
\end{figure}

To test the performances of our new algorithm we start by assessing its accuracy. We consider, for an assigned $M$, a random $\lambda \in [-0.999,0.999]$ and a random $x_{true} \in [-0.99,3]$ (or $x_{true} \in [-0.999,0.999]$ whenever $M>0$) and we compute the resulting time of flight $T$. We then use Housholder iterations starting from the appropriate initial guess to find back the $x$ value. We find that stopping the iterations whenever the difference between the $x$ value computed at two successive iterations is less than $10^{-5}$ ($10^{-8}$ whenever $M>0$) is a good setting. We record, for each of such trials, the number of iterations made by the Householder method $it$, and the error defined as $\epsilon = |x_{true}-x|$. This is repeated 1,000,000 times for $M=0$ and then 100,000 times for each $M=1,2,....,50$. The result is shown in Figure \ref{fig:accuracy}. In the vast majority of cases we obtain an error $<10^{-13}$. Few cases have a slightly larger error (up to $<10^{-11}$) and these are mainly corresponding to multirevolution cases where $T\approx T_{min}$.  We also note, on the proximity of $\lambda=1$ values, a distinct rise in the absolute error. This is due to the $M=0$ case and the loss of precision in the computation of $y$ from $\lambda$, a problem that can be avoided computing $y$ directly from the problem geometry, but it is here deemed as not necessary. Looking then at the number of iterations, we compute the mean over all instances having the same $M$ value. We obtain, for the single revolution case, an average of 2.1 iterations while, in the multiple revolution case, we get an average of 3.3 iterations to convergence. Note how in these tests we do not find a case where a switch occurs between the short period and long period solution during the root solving procedure. Such a switch would infact immediately appear in Figure \ref{fig:accuracy} as a point with a large absolute error $\epsilon$.

We then turn to the anaysis of our algorithm complexity with respect to the known Gooding algorithm, considered by many as the most accurate and efficient Lambert solver up to date. First we note that in terms of accuracy, Gooding algorithm is comparable to ours. We then run a speed test. For the purose of this test we reimplement both our and Gooding algorithm in pure Python language (i.e. no C++ bindings) and we record the execution time to solve the same 100,000 randomly generated problems (using the same bounds as above). In the case $M=0$, the proposed algorithm resulted to be faster by a factor $1.25$, while in the multi revolution cases by a factor $1.5$. This type of test is very sensitive to implementation details and to the underlying computing architecture and even if we did our best to pay as much attention to them in both cases, we support our result with more general considerations. The main difference between our algorithm and Gooding's is in the initial guess generation and in the iteration method. Gooding algorithm employs Halley iterations, while we make use of Householder iterative scheme. While Halley's method has a slighlty lower complexity and does not need to compute also the third derivative from Eq.(\ref{eq:derivatives}), our iterative scheme reaches, in the case $M=0$, a comparable accuracy in only 2 iterations on average compared to the 3 iterations needed for the Gooding case. For $M>0$ the number of required iterations is comparable in both cases but the initial Guess used in Gooding algorithm has, in general, a higher complexity as it makes use of a higher number of square roots and exponentiations. In the $M>0$ case Gooding initial guess also requires the determination of $x_{min},T_{min}$ via a further Halley iterative scheme, while the initial guess we use does not make use of any particular value, while still allowing the Householder method to converge within a few iterations and in all cases. We must, though, note once more that as we do not compute $x_{min},T_{min}$  we also cannot bound the solution during the root solver iterations and thus allow for the theoretical possibilty of a switch between short and long period solutions.  Such a rare event never appeared in our extensive testing of the new routine

Finally, we measure the error also in terms of the computed terminal velocities by comparing all $\mathbf v_2$, returned by our Lambert solver, to the same values as computed via numerical propagation (using Lagrange coefficients) from $\mathbf r_1,\mathbf v_1$. We do this by instantiating at random $\mathbf r_1, \mathbf r_2$ with each component in the range $[-4,4]$ and $t\in[0.1,100]$. For the purpose of this test we consider $\mu=1$ and we measure the norm of the resulting vector of the velocity difference. Repeating this experiment for a total cumulative 10,000,000 Lambert's Problems, an average error of $10^{-13}$ is obtained, with a maximum error measured to be $10^{-8}$.

\bibliographystyle{spbasic}      
\section{Conclusion}
We revisit Lambert's problem building upon the results of Lancaster and Blanchard and finding some new properties of the time of flight curves. We propose a new transformation of such curves able to further simplify the problem suggesting efficient approximations to the final solution. Using our results to design a new procedure to solve the Lambert problem we are able to build a low complexity algorithm that we find able to provide accurate solutions in a shorter time when compared to the state of the art Gooding's algorithm.

\bibliography{main}
\bibliographystyle{spbasic}

\end{document}